\documentclass[preprint2,numberedappendix]{emulateapj-rtx4}
\usepackage{natbib}
\usepackage{amsmath}
\usepackage{graphicx,times,bm,url}
\usepackage{afterpage,lscape}
\graphicspath{{./fig/}{./png/}}
\newcommand{\EQ}{\begin{equation}}
\newcommand{\EN}{\end{equation}}
\newcommand{\EQA}{\begin{eqnarray}}
\newcommand{\ENA}{\end{eqnarray}}

\newcommand{\Fig}[1]{Figure~\ref{#1}}

{}
{}
{}

{}
{}
{}
{}
{}
{}
{}
{}
{}
{}
{}
{}
{}
{}
{}
{}
{}
{}
{}
{}
{}

{}
{}
{}

{}

{}
{}

\newcommand{\xx}{\bm{x}}
\newcommand{\XX}{\bm{X}}

\newcommand{\BB}{\bm{B}}

\newcommand{\uu}{\mbox{\boldmath $u$} {}}

\newcommand{\JJ}{\mbox{\boldmath $J$} {}}

\newcommand{\AAA}{\mbox{\boldmath $A$} {}}

\newcommand{\ee}{\mbox{\boldmath $e$} {}}

\newcommand{\FF}{\mbox{\boldmath $F$} {}}

\newcommand{\nab}{\mbox{\boldmath $\nabla$} {}}

\newcommand{\eeps}{\mbox{\boldmath $\epsilon$} {}}
\newcommand{\sgn}{{\rm sgn}  \, {}}

\newcommand{\dd}{{\rm d} {}}

\usepackage{color}

\begin{document}

\title{Magnetic field relaxation and current sheets in an ideal plasma}

\author{S. Candelaresi}
\author{D. I. Pontin}
\author{G. Hornig}
\affil{
Division of Mathematics, University of Dundee, Dundee, DD1 4HN, UK
}

\begin{abstract}
We investigate the existence of magnetohydrostatic equilibria for topologically
complex magnetic fields.
The approach employed is to perform ideal numerical relaxation experiments.
We use a newly-developed Lagrangian relaxation scheme that exactly preserves the
magnetic field topology during the relaxation.
Our configurations include both twisted and sheared fields, of which some fall
into the category for which \cite{Parker-1972-174-499-ApJ} predicted no
force-free equilibrium.
The first class of field considered contains no magnetic null points,
and field lines connect between two perfectly conducting plates.
In these cases
we observe only resolved current layers of finite thickness.
In further numerical experiments we confirm that magnetic null points are
loci of singular currents.
\end{abstract}

\section{Introduction}

Magnetic field relaxation in environments like the solar atmosphere and
laboratory plasma is a crucial process in understanding
open problems like solar flares and field stability in tokamaks.
In such environments the field evolves nearly ideally, i.e.\ the magnetic flux
remains frozen to the plasma.
For an arbitrary braided magnetic field between two perfectly conducting planes
\cite{Parker-1972-174-499-ApJ} hypothesized that there can be a force-free
equilibrium of the same topology only if the field's twist varies
uniformly along the large-scale magnetic field.
He further suggested that in resistive magnetohydrodynamics (MHD), where
reconnection can occur, the field would then undergo a rapid change in
topology accompanied by magnetic energy dissipation that would provide a
significant contribution to coronal heating \citep{Parker-1983-264-635-ApJ}.

In subsequent works this idea has been confirmed and challenged
various times \citep{Parker-1983-264-635-ApJ, Craig-Sneyd-2005-232-1-SolPhys,
Low-2010-266-2-SolPhys, Low-2013-768-7-ApJ}.
Braided magnetic fields from foot point motions were shown to be complex
enough that they must exhibit the proposed topological dissipation
\citep{Parker-1983-264-642-ApJ}.
\cite{Low-2010-266-2-SolPhys} later showed that there exist solutions for the
relaxing magnetic field which permit current sheets.
One of the first simulations testing the conjecture was
performed by \cite{Mikic-Schnack-1989-338-1148-ApJ} who found filamentary
current structures with an exponentially increasing strength.
Given the limited computing power of that time, they were only able to
reach very moderate resolutions, which renders it questionable if they
observed proper sheets.

Doubts about Parker's conjecture came from e.g.\
\cite{Ballegooijen-1985-298-421-ApJ} who suggested that a field generated
by foot point motions is able to adjust to those motions and reach a force-free
state so long as the velocity field is continuous at the boundary.
This was supported by later numerical simulations, in which a series of
footpoint displacements were performed, and an exponential thinning and
intensification of current layers was observed -- rather than a collapse to
sub-grid scale of the current \citep{vanballegooijen1988a}.
It has also been suggested that in certain configurations no thin current
layers -- finite or infinite -- need necessarily form.
\cite{Craig-Sneyd-2005-232-1-SolPhys} derived solutions for relaxing
magnetic fields which do not show singularities even with sufficiently braided
configurations.
However, \cite{pontin2015} recently demonstrated that for any braided magnetic
field in which the field line mapping exhibits small length scales, thin current layers
are an inevitable feature of the corresponding force-free equilibrium, if it exists.
Building on earlier work by \cite{Wilmot-Smith-Hornig-2009-696-1339-ApJ}
they showed that the ideal relaxation of a class of braided fields leads to a
current distribution of finite strength. Moreover, the current layers obtained in the
approximate force-free equilibria were shown to scale in both thickness and
intensity with length scales present in the field line mapping, consistent with
the earlier results of \cite{vanballegooijen1988a}.

In this work we tackle the problem of current sheet formation during magnetic field
relaxation for various topologically non-trivial configurations at unprecedented
numerical resolution.
\cite{longcope1994} pointed out that there exist solutions for relaxed magnetic
fields which have current layers thin enough that they cannot be distinguished
from current sheets with moderate grid resolution.
We apply the newly developed numerical code {\textsc GLEMuR}
\citep{Candelaresi-2014-mimetic} which uses the resources of graphical
processing units (GPUs) and makes use of mimetic differential operators
\citep{Hyman-Shashkov-1997-33-4-CompAthApp}, which greatly improve the
relaxation quality.
The scheme is Lagrangian, and is constructed in such a way that it perfectly preserves the magnetic topology \citep{Craig-Sneyd-1986-311-451-ApJ}.

Emphasis is put on braids which are not reducible to uniform twists along a
mean magnetic field such as those used by
\cite{Wilmot-Smith-Hornig-2009-696-1339-ApJ}, as well as fields generated through
footpoint motions such as those by \cite{Longbottom-Rickard-1998-500-471-ApJ}.
We further investigate the effect of modifying the magnetic field to include magnetic
null points, and show that current singularities form there
\citep[as in][]{pontincraig2005, Craig-Pontin-2014-788-2-ApJ}.

\section{Model and Methods}\label{methodsec}

\subsection{Ideal Evolution}
In order to determine existence and structure of equilibria for given magnetic
topologies, we require to follow an exactly ideal evolution. We employ a method
that by its construction exactly preserves the magnetic flux, magnetic field line
connectivity, and solenoidal nature of the magnetic field $\BB$ during the relaxation.
Specifically, we use the Lagrangian code {\textsc GLEMuR}
\citep{Candelaresi-2014-mimetic} which solves the equations for an
ideal evolution of a magnetized non-Newtonian fluid without inertia, as well
as an extension to this method that considers a damped fluid with inertia.
These methods have computational advantages over those that solve for
the full dynamics of ideal MHD, leading towards a minimum energy state
whose properties are our main concern (rather than the evolution to reach the
relaxed state).

In order to preserve the field's topology we  make use of a Lagrangian
grid method where the grid points move along with the fluid.
If the initial positions of fluid particles at time $t = 0$ are described by
the position vector field $\XX$, we
denote their position at time $t$ by $\xx(\XX, t)$ with $\xx(\XX, 0) = \XX$.
These fluid elements (grid points) are evolved according to
\EQ
\frac{\partial \xx(\XX,t)}{\partial t} = \uu(\xx(\XX,t),t),
\EN
where the velocity $\uu$ is chosen in such a way to lead towards an equilibrium.
We employ different methods for choosing $\uu$, as outlined below.

Any ideal evolution of the magnetic field $\BB$ must be consistent with the
ideal induction equation
\EQ
\frac{\partial \BB}{\partial t} - \nab\times(\uu\times\BB) = 0,
\EN
which implies that the magnetic field is frozen into
the fluid \citep{BatchelorFrozeIn1950RSPSA,PriestReconnection2000},
i.e.\ moves together with the fluid particles.
From the frozen in condition
we can relate the magnetic field at later time (following a deformation
of the fluid particle mesh) to the magnetic field at $t = 0$;
\EQ
B_{i}(\XX,t) = \frac{1}{\Delta} \sum_{j=1}^{3} \frac{\partial x_{i}}{\partial X_{j}}
B_{j}(\XX,0),
\EN
with $B_i$ being the $i^{\rm th}$ component of the magnetic field and
$\Delta = \det(\partial x_{i}/\partial X_{j})$ \citep{Craig-Sneyd-1986-311-451-ApJ,Candelaresi-2014-mimetic}.
Here the fields are functions of their initial positions $\XX$ and time $t$.
In other words, they are functions of the fluid particle positions.

For some of the relaxation simulations described herein, we follow
\cite{Candelaresi-2014-mimetic} by applying the magneto-frictional
term \citep{Chodura1981} for the evolution of the fluid
\EQ \label{eq: mf}
\uu = \JJ\times\BB,
\EN
with the current density $\JJ = \nab\times\BB$.
This is the evolution equation for a non-Newtonian fluid without inertia, and
the evolution terminates when a force-free field (satisfying $\JJ\times\BB={\bf 0}$)
is attained.
This approach is well suited
for studying relaxation problems, as it is shown to
lead to a monotonic decay of the magnetic energy
\citep{Craig-Sneyd-1986-311-451-ApJ,Yang-Sturrock-1986-309-383-ApJ}.

However, there are two disadvantages to this approach.
First, the monotonic energy
decay means that during the relaxation the system is unable to escape any small
local energy minima if a lower global energy minimum exists.
Second, in a magnetic field
containing null points, the null point positions are fixed (since the $\JJ\times\BB$ force
at the nulls themselves must be zero).
To address the first issue we consider an extension of the method that makes use of inertial effects.
The fluid's evolution equation is then given by
\EQ \label{eq: inertial}
\frac{\dd\uu}{\dd t} = (\JJ\times\BB - \nu\uu)/\rho,
\EN
with the damping coefficient $\nu$ and density $\rho$.

To address the second issue of stationary
magnetic null points we employ a pressure force.
In some cases described below it is beneficial to seek an equilibrium that is
not force-free, but where the Lorentz force is balanced by a pressure gradient.
For simplicity here we assume that the pressure is directly proportional to the
fluid density (corresponding to an ideal gas under isothermal changes of state).
This yields an evolution of the fluid mesh
\EQ \label{eq: mf gradP}
\uu = \JJ\times\BB - \beta\nab\rho,
\EN
with the compressibility parameter $\beta$.
The density can be expressed in terms
of the initial density $\rho_0$ as $\rho(\xx,t) = \rho_{0}/\Delta = \rho(\XX,0)/\Delta$, and
for convenience we will always choose $\rho_{0} = 1$.
We can also add the pressure gradient to the inertial evolution equation,
to give
\EQ \label{eq: inertial gradP}
\frac{\dd\uu}{\dd t} = (\JJ\times\BB - \nu\uu - \beta\nab\rho)/\rho.
\EN

Computing spatial derivatives on a moving grid is a sensitive operation.
The direct approach used in previous numerical implementations of the
magneto-frictional approach involves application of the chain rule leading to
expressions involving various products of derivatives
\citep{Craig-Sneyd-1986-311-451-ApJ}.
Using such direct derivatives for computing $\JJ = \nab\times\BB$ on highly
distorted grids, such as those we expect to occur in our numerical experiments,
leads to numerical inaccuracies, most notably the issue that
$\nab\cdot\JJ = 0$ is not well fulfilled, as was noted by
\cite{Pontin-Hornig-2009-700-2-ApJ}.
Our code GLEMuR makes use of mimetic numerical operators to compute
the curl, which have been shown to more accurately represent the
current on such meshes, and have the advantage that they preserve the
identity $\nab\cdot(\nab\times\BB) = 0$ up to machine precision for some
appropriate mimetic divergence operator
\citep{Hyman-Shashkov-1997-33-4-CompAthApp, Candelaresi-2014-mimetic}.
For the time stepping we use a Runge-Kutta 6th order in time approach.

All three boundary conditions can be chosen to be periodic or line-tied.
Here line-tied means that the velocity is set to zero and the normal
component of the magnetic field is fixed.
For studying the problem proposed by \cite{Parker-1972-174-499-ApJ}
we will typically use such line-tied boundaries in the $z$ direction
in the simulations described below.
But occasionally we will impose periodic boundaries.

\subsection{Diagnostic Parameters}

Here we describe some diagnostic tools that are used in the following sections
to analyse the properties of the final states of our relaxation simulations.
The evolution of the system by equation \eqref{eq: mf}
is solely determined by the Lorentz force
$\FF_{\rm L} = \JJ\times\BB$.
A force-free state implies $\FF_{\rm L} = 0$, which is equivalent
to $\nab\times\BB = \alpha\BB$, where $\alpha$ is the force-free parameter
which satisfies $\nab\alpha\cdot\BB = 0$, i.e.\ $\alpha$ is constant along
magnetic field lines.
During the relaxation simulations, the magnetic field evolves into an energetically
more favorable state with approximately vanishing Lorentz force (when $\beta = 0$).
Since the Lorentz force never vanishes identically in this numerical approximation, the condition
$\nab\times\BB = \alpha\BB$ is not fulfilled exactly either.
We can, nevertheless, still  express the curl of the magnetic field in terms of
a component parallel and perpendicular to $\BB$:
\EQ \label{eq: non ff field}
\nab\times\BB = \lambda\BB - \eeps\times\BB,
\EN
with the parameter $\lambda$ and vector $\eeps$, where we choose $\eeps$ such
that $\eeps\cdot\BB = 0$.
These two parameters are used to determine the deviation from the force-free
state quantitatively.

From equation \eqref{eq: non ff field} we obtain
\EQ
\lambda = \frac{\JJ\cdot\BB}{\BB^{2}},
\EN
\EQ
\eeps = \frac{\JJ\times\BB}{\BB^{2}}.
\EN
Comparing $\lambda$ and $\eeps$ for each field line we can infer to what
degree the field is force-free.
For that we need to trace magnetic field lines from the bottom of the domain
at $z = Z_{0}$ to the top at $z = Z_{1}$ and integrate $\lambda$ and
$|\eeps|$ along the lines $C$:
\EQ
\lambda(X,Y) = \frac{1}{L} \int\limits_{C} \frac{\JJ\cdot\BB}{\BB^{2}}\ \dd l,
\EN
\EQ
\epsilon(X,Y) = \frac{1}{L} \int\limits_{C} \left|\frac{\JJ\times\BB}{\BB^{2}}\right|\ \dd l,
\EN
\EQ
L = \int\limits_{C} \dd l,
\EN
where we start our field line integration at $(X,Y,Z_{0})$.
The ratio $\epsilon(X,Y)/\lambda(X,Y)$ gives the relative deviation
from the force-free state.
Since $\BB\cdot\nab\lambda = 0$ for the force-free state we also compute the
maximum slope of $\lambda$ along the field lines in analogy to
\cite{Pontin-Hornig-2009-700-2-ApJ} and \cite{Candelaresi-2014-mimetic}:
\EQ
\lambda^{*}(X,Y) = \max_{i}\left(\frac{\lambda_{i+1}-\lambda_{i}}
{l_{i+1}-l_{i}}\right),
\EN
with the value $\lambda_{i}$ at point $i$ on the field line
and the length of the field line $l$.

For magnetic field lines extending between two parallel planes
\cite{Berger-1986-34-265-GAFD} suggested a relation between magnetic
helicity density and the winding of the field around itself.
Because the magnetic helicity density $h_{\rm m}$ is defined via the magnetic
vector potential $\AAA$ we choose to measure the twist of the magnetic field
lines by
\EQ
\omega(X,Y) = \frac{1}{L} \int\limits_{C} \frac{\JJ\cdot\BB}{|\JJ||\BB|} \ \dd l.
\EN
For a force-free field this expression reduces to $\omega(X,Y) = \sgn(\alpha)$.

\cite{Wilmot-Smith-Pontin-2010-516-A5-AA} showed that magnetic field lines
with a high integrated electric current are places of current sheet formation
and hence reconnection.
In our ideal simulations no reconnection can occur, but of course the formation
of localized current concentrations may take place.
To analyze their occurrence we compute the magnetic field line integrated current
density
\EQ \label{eq: Jp}
\JJ_{||}(X,Y) = \int\limits_{C} \frac{\JJ\cdot\BB}{|\BB|} \ \dd l.
\EN

\section{Braided Fields}\label{braidsec}

From previous numerical experiments \citep{Wilmot-Smith-Hornig-2009-696-1339-ApJ}
we know that
topologically complex braids do not necessarily form singular current sheets
as the field relaxes towards a force-free state.
Here we investigate the relaxation behavior of the magnetic braids discussed
by e.g.\ \cite{Wilmot-Smith-Hornig-2009-696-1339-ApJ} and \cite{Yeates_Topology_2010}.
To study the relaxation of these fields we use the magneto-frictional
evolution given by Equation (\ref{eq: mf}).

\begin{figure}[t!]\begin{center}
\includegraphics[width=0.95\columnwidth]{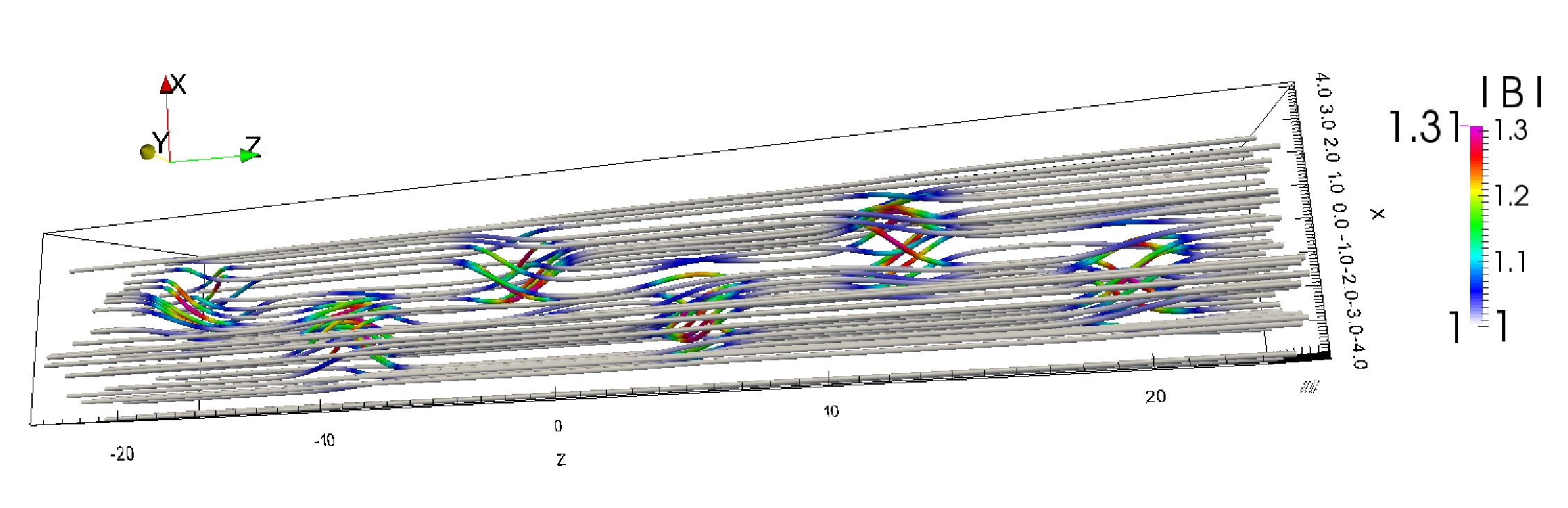}
\end{center}
\caption[]{
Initial magnetic field lines for the $E^3$ configuration.
Colors denote the field strength which is strongest in the twist regions.
}
\label{fig: e3_65_eps0_zUp1_p32_bfield_t0}\end{figure}

The initial magnetic field we consider is the one named $E^3$ by
\cite{Wilmot-Smith-Hornig-2009-696-1339-ApJ}, which consists of three
braiding regions and a homogeneous background magnetic field such that
$B_z > 0$ everywhere.
Its form is given by
\begin{eqnarray}
\BB_{\rm E^3}(0) & = B_0\ee_z+\sum_{c = 1}^{6} \frac{2kB_{0}}{a} (-(y-y_{c})\ee_x + (x-x_{c})\ee_{y}) & \nonumber \\
 & \times \exp\left(\frac{-(x-x_{c})^2-(y-y_{c})^2}{a^2} - \frac{(z-z_{c})^2}{l^2}\right), \label{eq:e3}&
\end{eqnarray}
with the initial field strength $B_{0}$, strength of twist $k$, radius
and length in $z$-direction of the twist region $a$ and $l$ respectively
and the twist locations $(x_c, y_c, z_c)$.
We choose $x_c = \{1,-1,1,-1,1,-1\}$, $y_c = 0$, $z_c = \{-20,-12,-4,4,12,20\}$,
$a = \sqrt{2}$, $l = 2$ and $B_{0} = 1$.
To fit this configuration into a computational domain, the box size is chosen to extend
8 units in $x$ and $y$ and 48 units in $z$ centered at the origin.
Upper and lower boundaries are chosen either to be line-tied or periodic
and the grid resolution is $300$ in each direction.
Sample magnetic field lines are shown in \Fig{fig: e3_65_eps0_zUp1_p32_bfield_t0}.

\subsection{Formation of Current Layers}

\begin{figure}[t!]\begin{center}
\includegraphics[width=0.95\columnwidth]{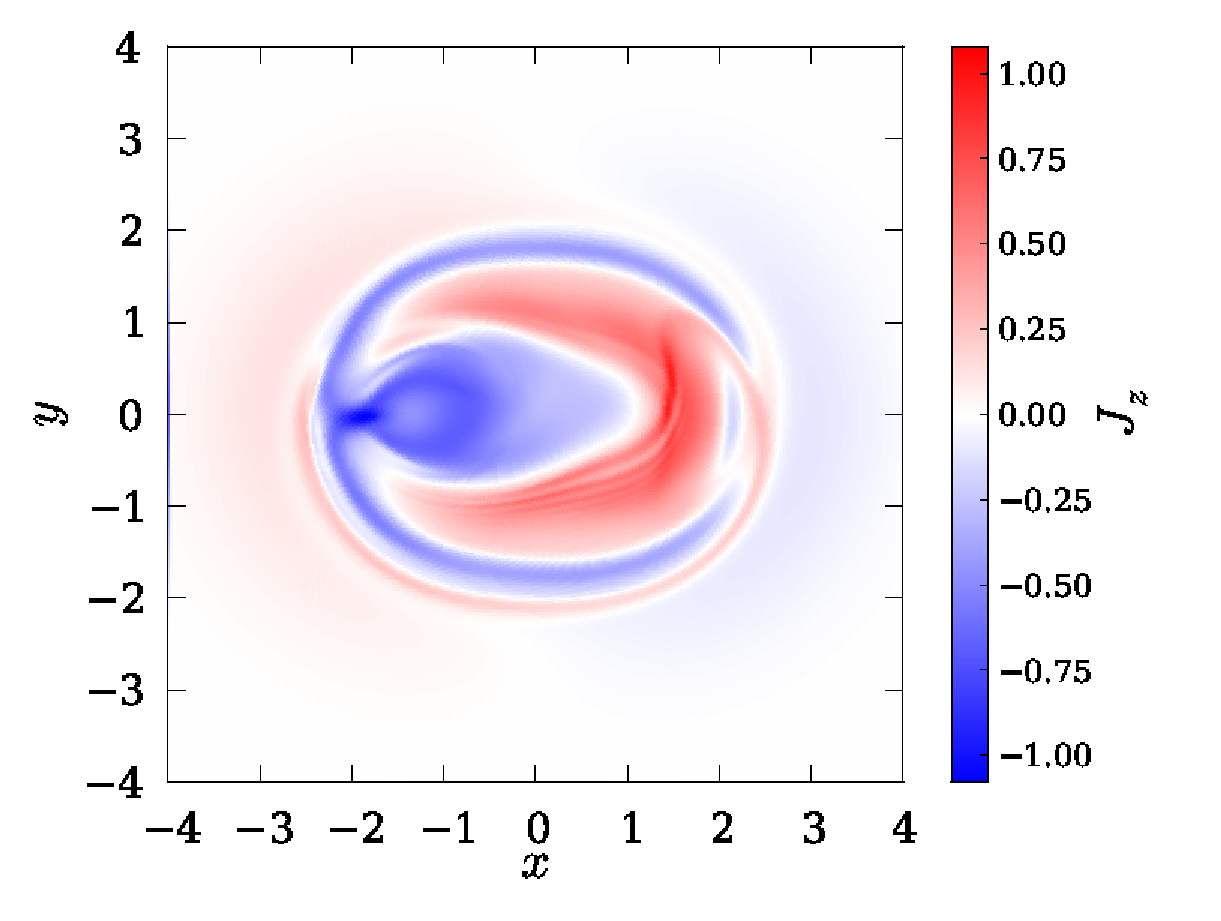} \\
\includegraphics[width=0.95\columnwidth]{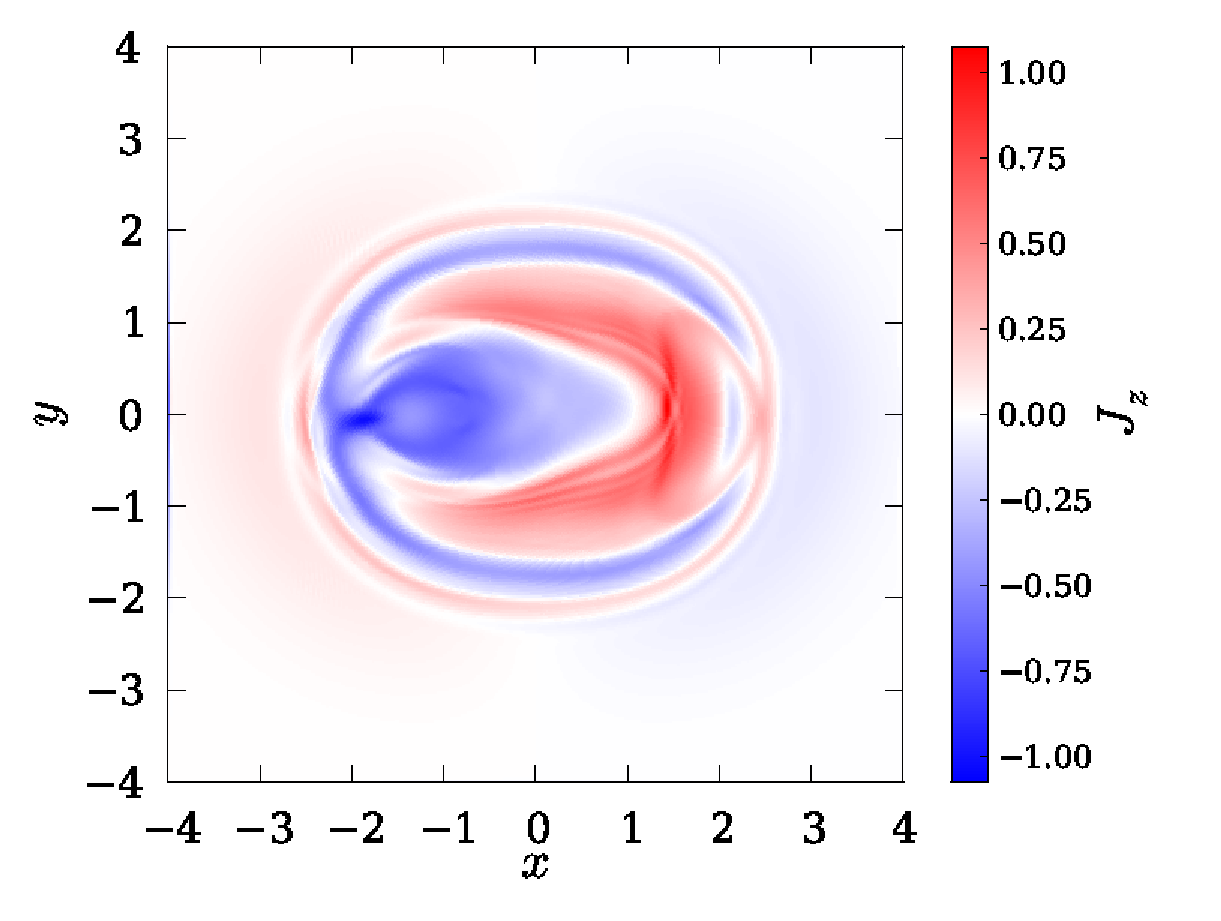}
\end{center}
\caption[]{
$z$-component of the electric current density at $z = 3.8$ for times close to
relaxation for $E^3$ with
line tied (upper panel) and periodic boundaries in z (lower panel).
}
\label{fig: braids current}\end{figure}

As the field evolves and tries to minimize the magnetic energy it forms
concentrations of strong currents.
According to \cite{Parker-1972-174-499-ApJ} singular current sheets should
form.
However, we do not find any such formation irrespective of the grid resolution
(\Fig{fig: braids current}, upper panel) and all current concentrations
are well resolved
which favors Ballegooijen's result \citep{Ballegooijen-1985-298-421-ApJ}.
This is even true if we choose periodic boundaries in z-direction
(\Fig{fig: braids current}, lower panel).

Varying the grid resolution does not significantly change the outcome of
these simulations.
The width of the current layers remains the same, as well as the strength
of the current.

\subsection{Topological Complexity}

\begin{figure}[t!]\begin{center}
\includegraphics[width=0.95\columnwidth]{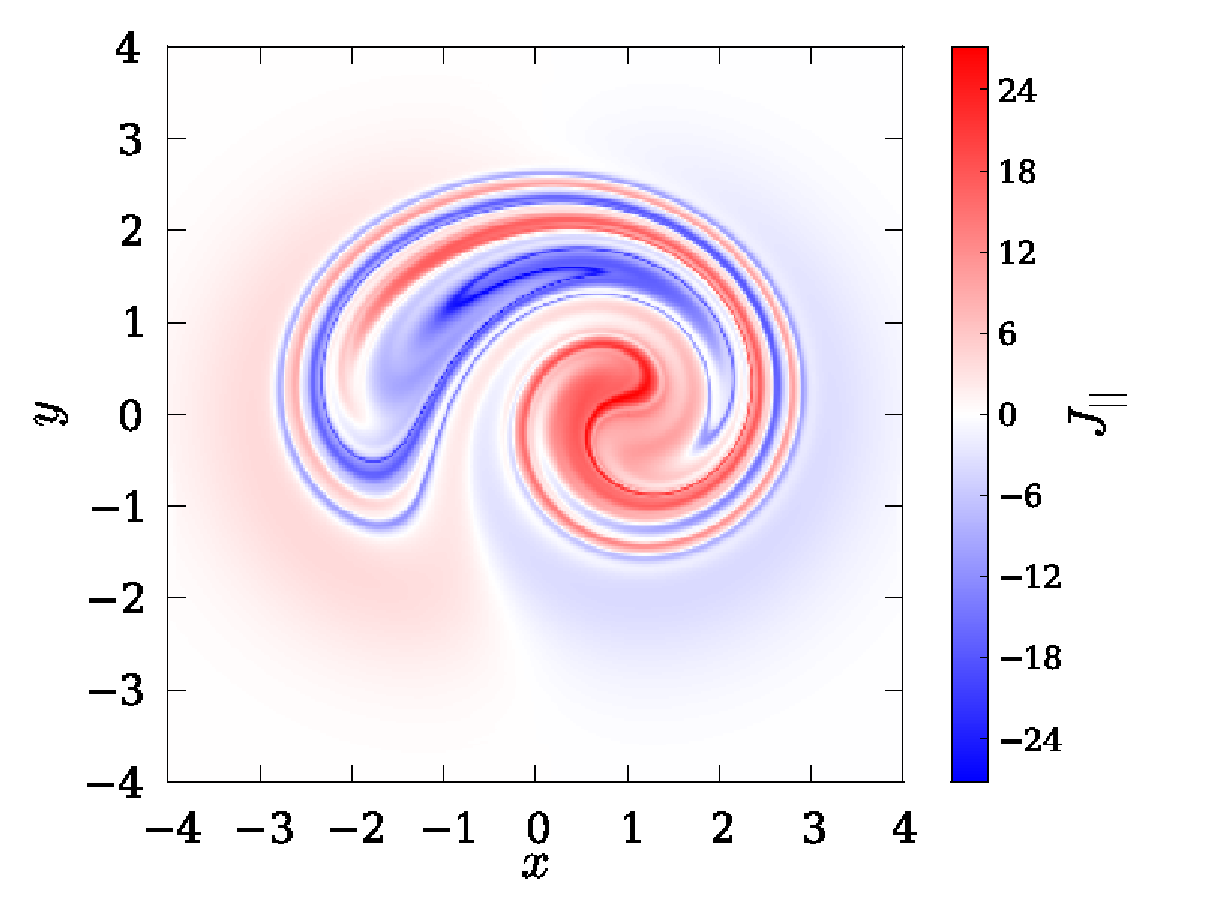} \\
\includegraphics[width=0.95\columnwidth]{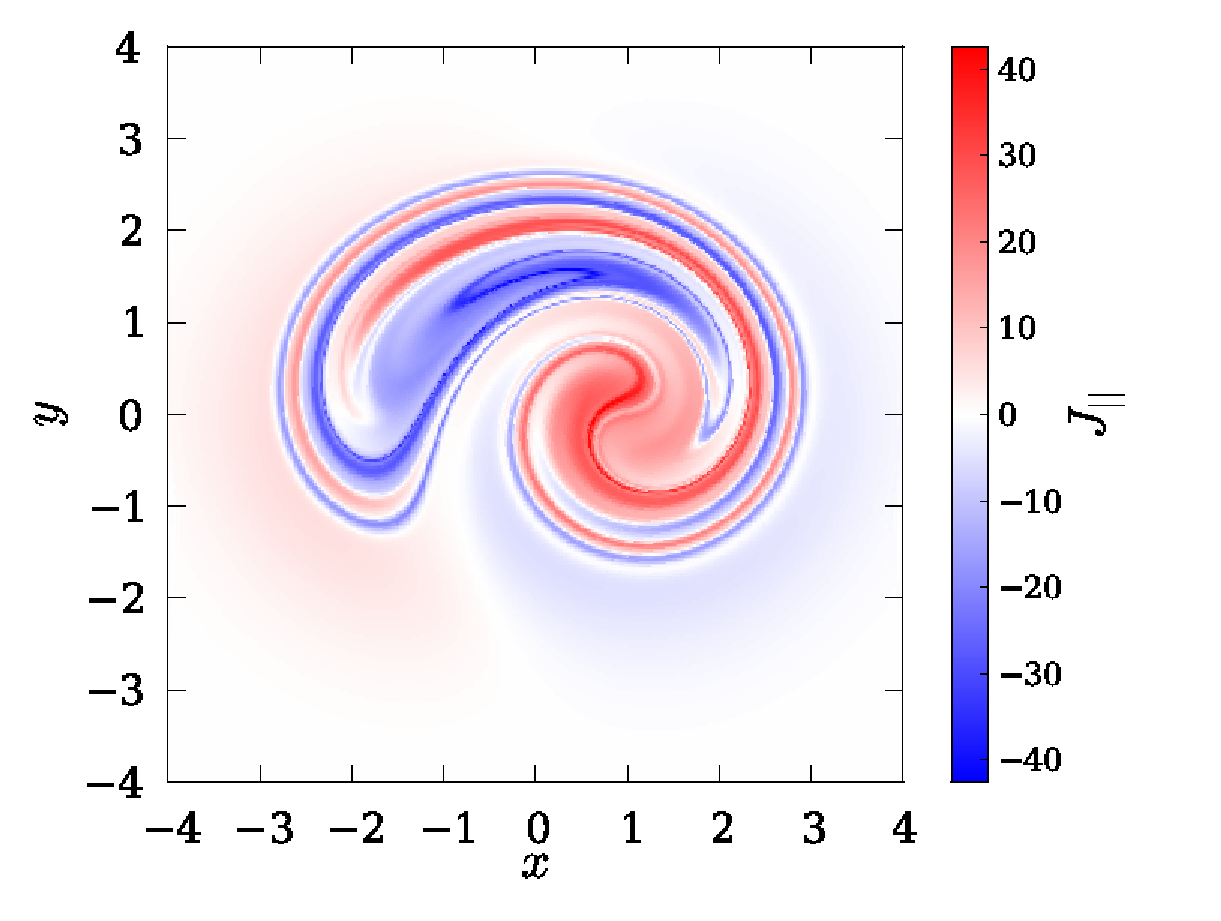}
\end{center}
\caption[]{
Line integrated electric current density $J_{||}(X,Y)$ as computed from
equation \eqref{eq: Jp} for $E^3$ with
line tied boundaries at initial time $t = 0$ and time at relaxation
$t = 60$.
}
\label{fig: braids Jp}\end{figure}

Since the evolution of the magnetic field is ideal it preserves its topology
and changes in connectivity are forbidden.
One measure of the field's topological complexity is the field line integrated
electric current density $\JJ_{||}(X,Y)$.
We observe an approximate conservation for $\JJ_{||}$ for our test configuration
of $E^3$ (\Fig{fig: braids Jp}).
From \Fig{fig: braids Jp} it is readily seen  that despite the simple structure
of $J_z$ in \Fig{fig: braids current}
the thin structure of $J_{||}$ demonstrates the high complexity of the
field line configuration $E^3$.

\subsection{Force-Freeness}

Whether or not arbitrarily twisted flux concentrations are allowed to evolve
into a force-free state is the second aspect of Parker's conjecture.
Here we monitor the evolution of the force-free parameter $\lambda^{*}$,
line averaged Lorentz force $\epsilon$
and the twist $\omega$ for all field lines.

\begin{figure}[t!]\begin{center}
\includegraphics[width=0.95\columnwidth]{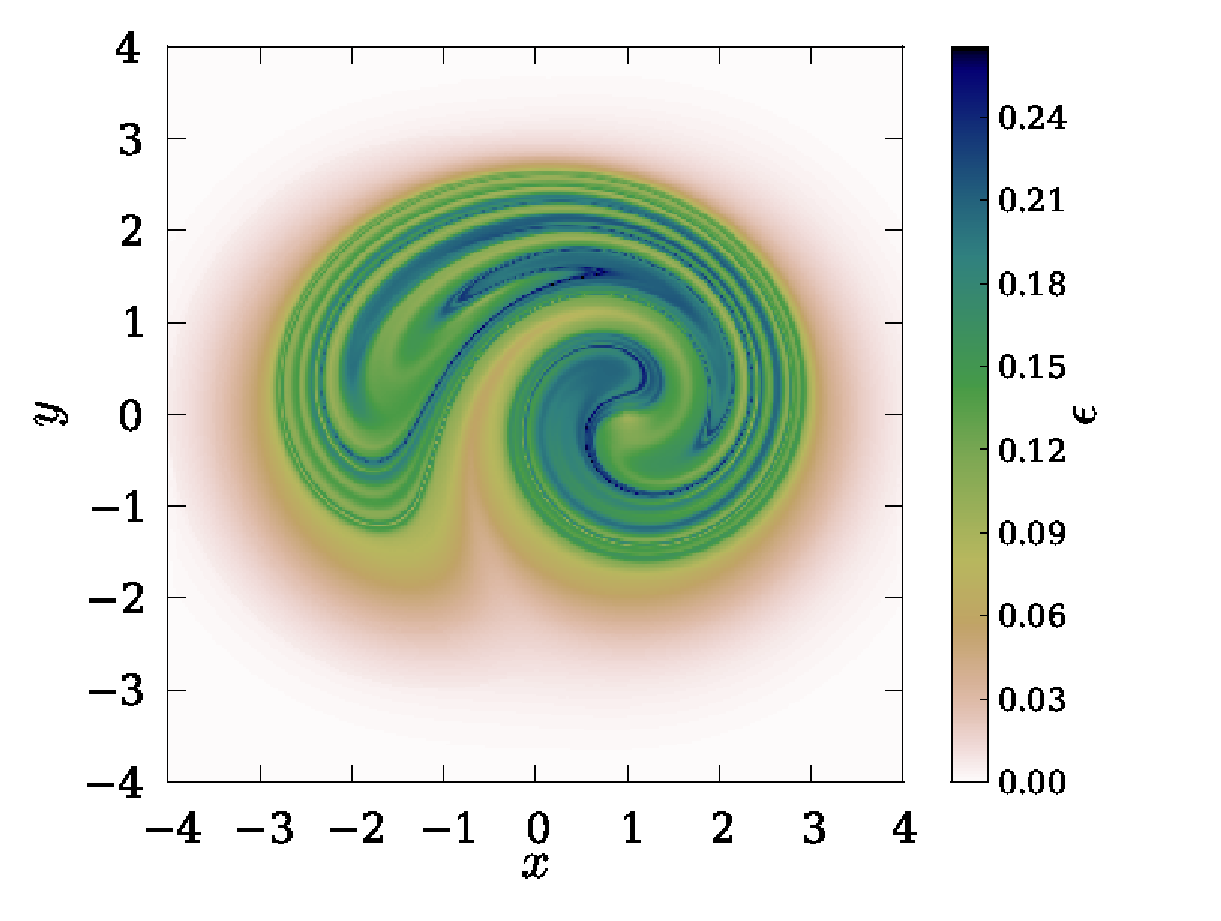} \\
\includegraphics[width=0.95\columnwidth]{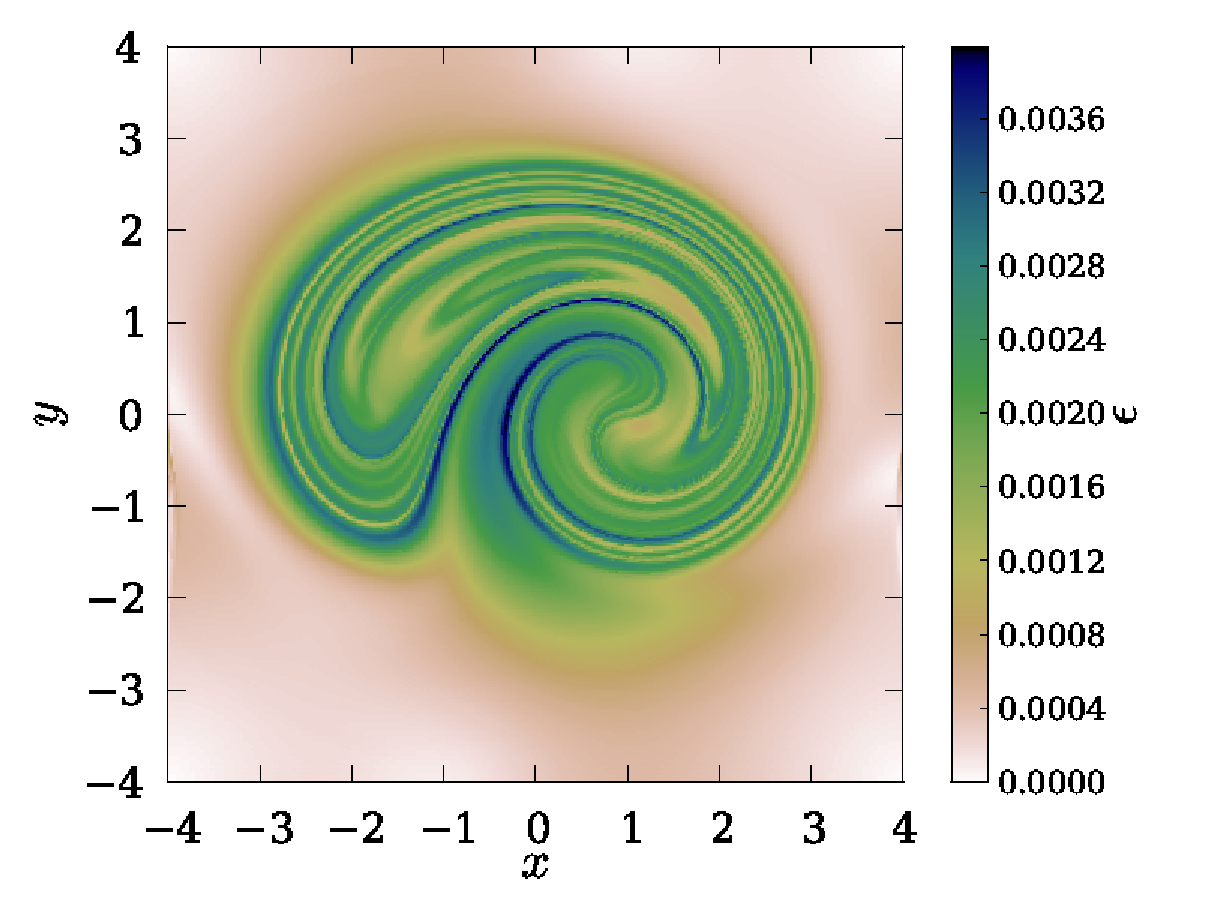}
\end{center}
\caption[]{
Average modulus of the Lorentz force along magnetic field lines
for the $E^3$ configuration with line tied boundaries
at $t = 0$ (upper panel) and $t = 60$ (lower panel).
}
\label{fig: braids epsilon}\end{figure}

In line with previous simulations by \cite{Craig-Sneyd-1986-311-451-ApJ},
\cite{Pontin-Hornig-2009-700-2-ApJ} and \cite{Candelaresi-2014-mimetic}
the field evolves such that the domain maximum and average of the Lorentz
force decreases in time (\Fig{fig: braids epsilon}).
This decrease is, however, not uniform in the field lines.
While $\epsilon$ is rather smooth at the beginning, it develops large
gradients and small-scale structures as the field relaxes.
In those thin loops the Lorentz force no longer decreases and prevents
the whole system from reaching a force-free equilibrium.

\begin{figure}[t!]\begin{center}
\includegraphics[width=0.95\columnwidth]{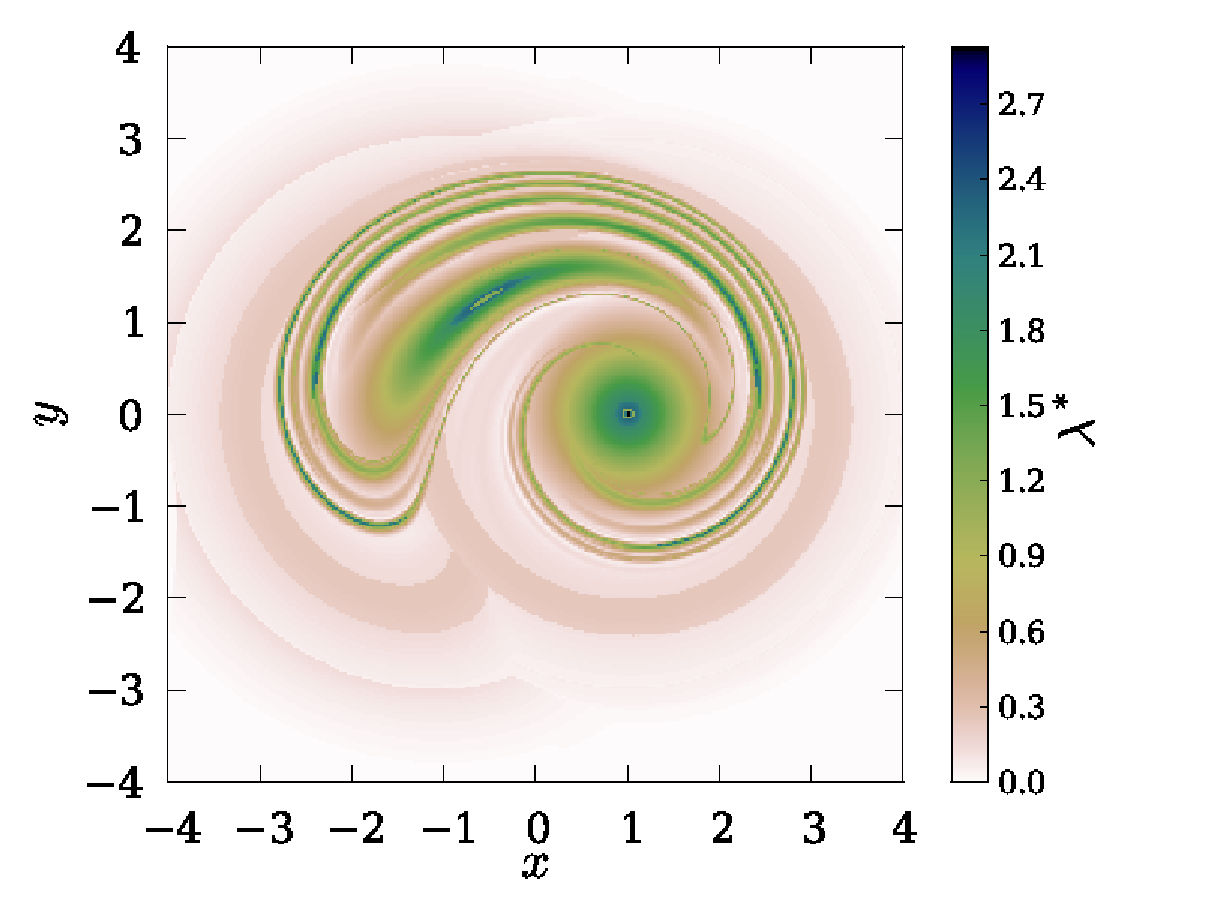} \\
\includegraphics[width=0.95\columnwidth]{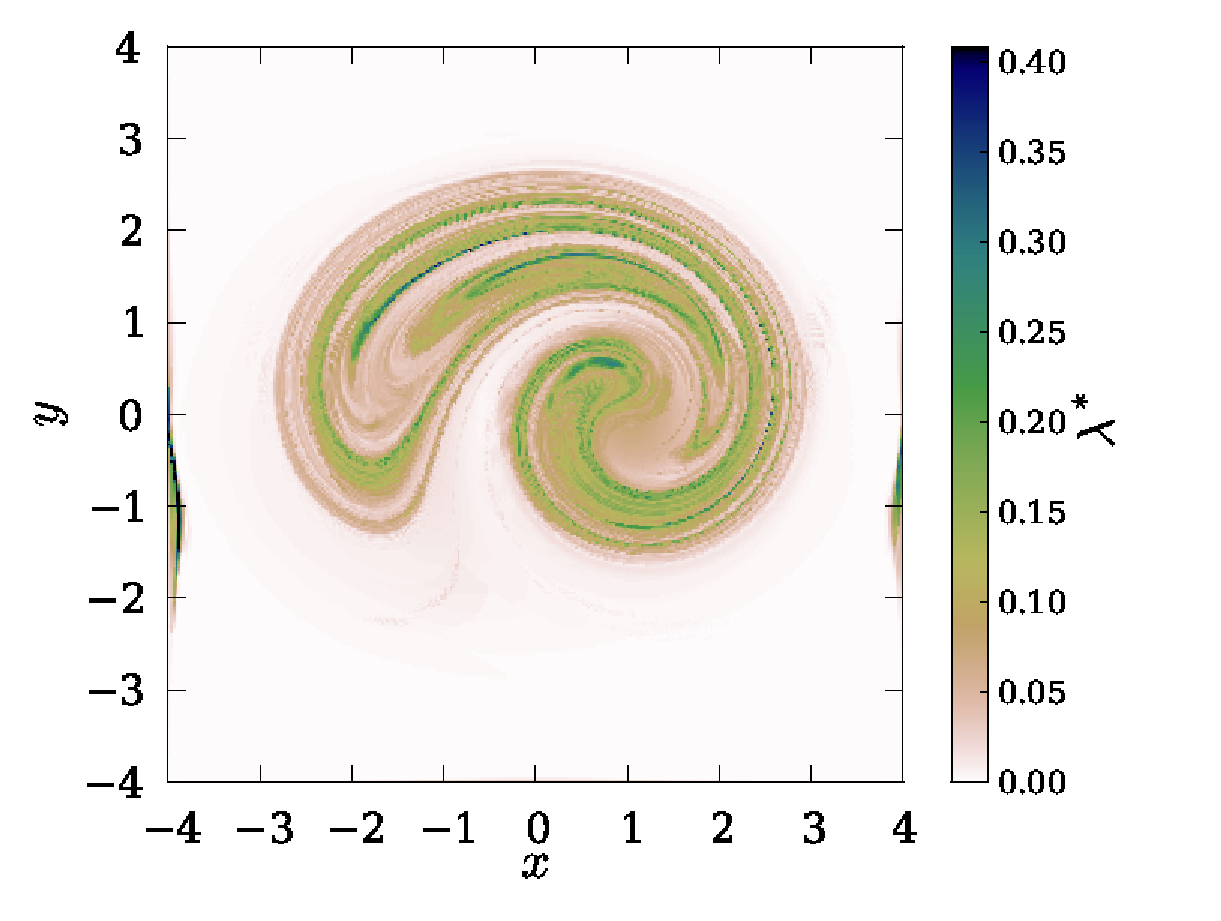}
\end{center}
\caption[]{
Maximum gradient of the force-free parameter $\lambda$ along all magnetic field lines
for the $E^3$ configuration with line tied boundaries
$t = 0$ (upper panel) and $t = 60$ (lower panel).
}
\label{fig: braids lambdaS}\end{figure}

While $\epsilon$ measures the strength of the forces  along the field lines,
$\lambda^{*}$ measures the deviation from the force-free state, i.e.\
$\nab\times\BB = \alpha\BB$ with $\BB\cdot\nab\alpha = 0$.
% Since $\lambda^{*}$ has the unit $1/{\rm length}^2$ we compare it with the
% typical scale of the field's variation in $z$-direction which corresponds
% to the parameter $l$ from equation \eqref{eq:e3}.
% We find $\max_{(X,Y)}\lambda^{*}l^2 = 1.6$ which means a moderate variation
% along the field line.
As expected, the system approaches a state close to force-free
(\Fig{fig: braids lambdaS}).
At the same time it develops small-scale features, like $\epsilon$
where $\lambda^{*}$ does not change significantly.
Those features are a characteristic of this highly twisted field
which were illustrated by e.g.\ \cite{Yeates_Topology_2010}.
From \Fig{fig: braids lambdaS} we can conclude that, although
the overall system approaches a more force-free state it does so only
on average while locally being prevented to reach that state.

\begin{figure}[t!]\begin{center}
\includegraphics[width=0.95\columnwidth]{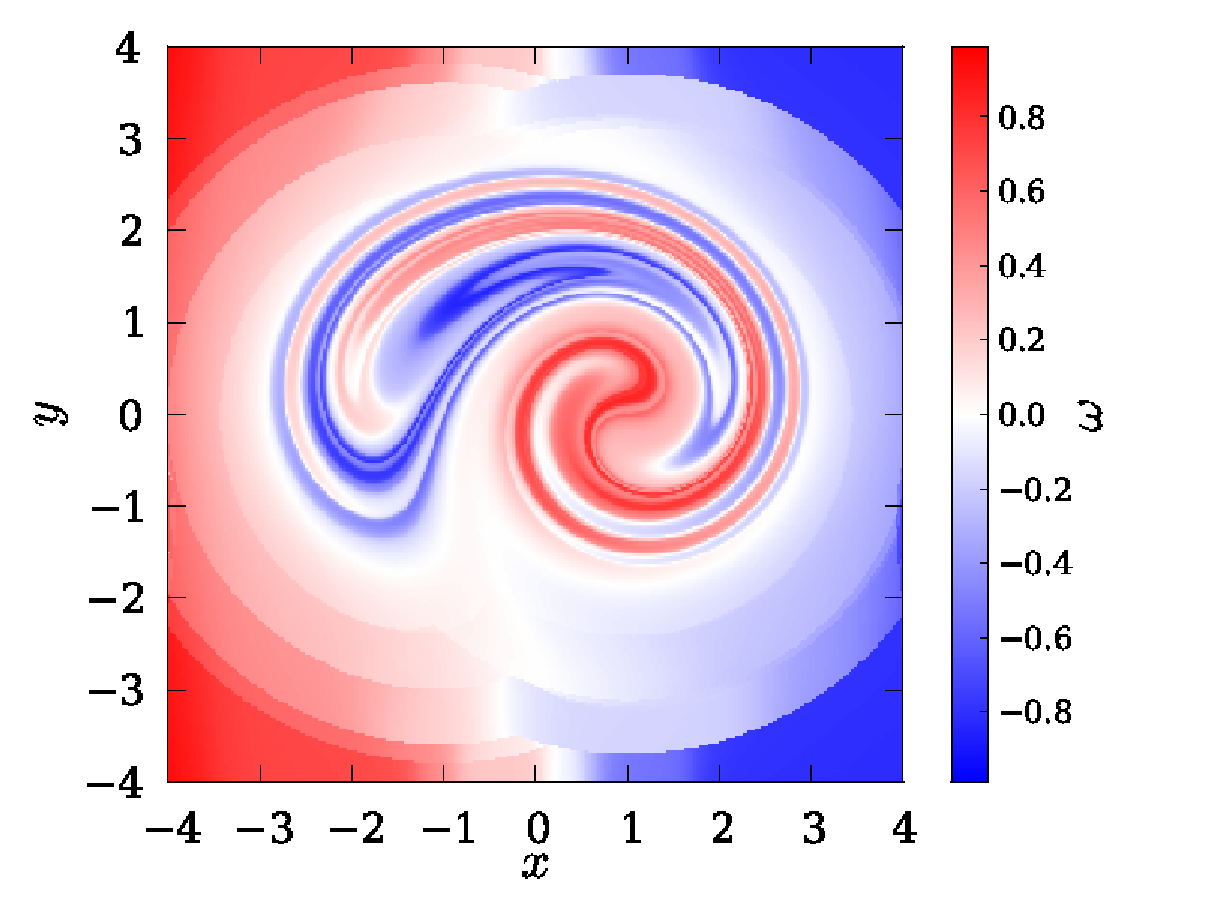} \\
\includegraphics[width=0.95\columnwidth]{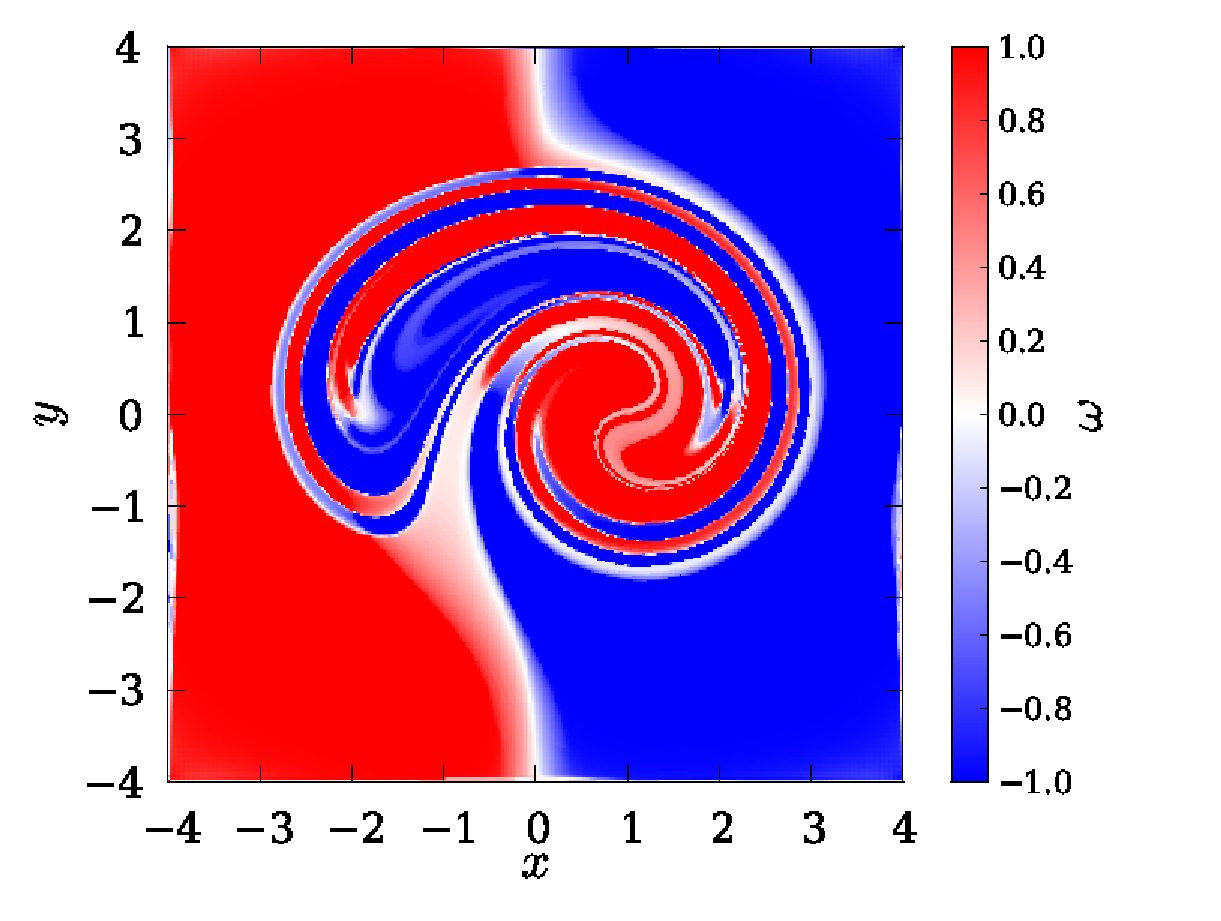}
\end{center}
\caption[]{
Average twist of the magnetic field lines $\omega$
for the $E^3$ configuration at $t = 0$ (upper panel) and $t = 60$ (lower panel).
}
\label{fig: braids omega}\end{figure}

By using color maps of the magnetic field line \cite{Yeates_Topology_2010}
showed that regions of different field line mappings are connected to a
non-trivial topology of the field.
Similarly we observe regions where the sign of the twist
$\omega$ changes sharply (\Fig{fig: braids omega}).
Those are exactly the loci where both $\epsilon$ and $\lambda^{*}$ develop
into thin structures and $\lambda^{*}$ stays approximately constant in time.

The reason that $\epsilon$ and $\lambda^{*}$ develop thin structures as the
relaxation proceeds is not clear.
This could be a feature of the numerical
method employed to perform the relaxation: specifically that under certain conditions
the scheme acts to reduce the $\JJ\times\BB$ force on average within the domain
at the expense of particular locations at which the relaxation is compromised.
On the other hand, it is possible that this is associated with some more fundamental
property of the magnetic field.
In particular, it could be that the topology of the field, as manifested through the
sign change of the
average field line twist $\omega$ impedes the further evolution
of the field into a perfectly force-free state.
In order to determine whether this
is the case we require to develop a theory for the evolution of these quantities.
This is outside the scope of the present study.

\section{Current Formation at Null-Points}\label{nullsec}

\begin{figure}[t!]\begin{center}
\includegraphics[width=0.95\columnwidth]{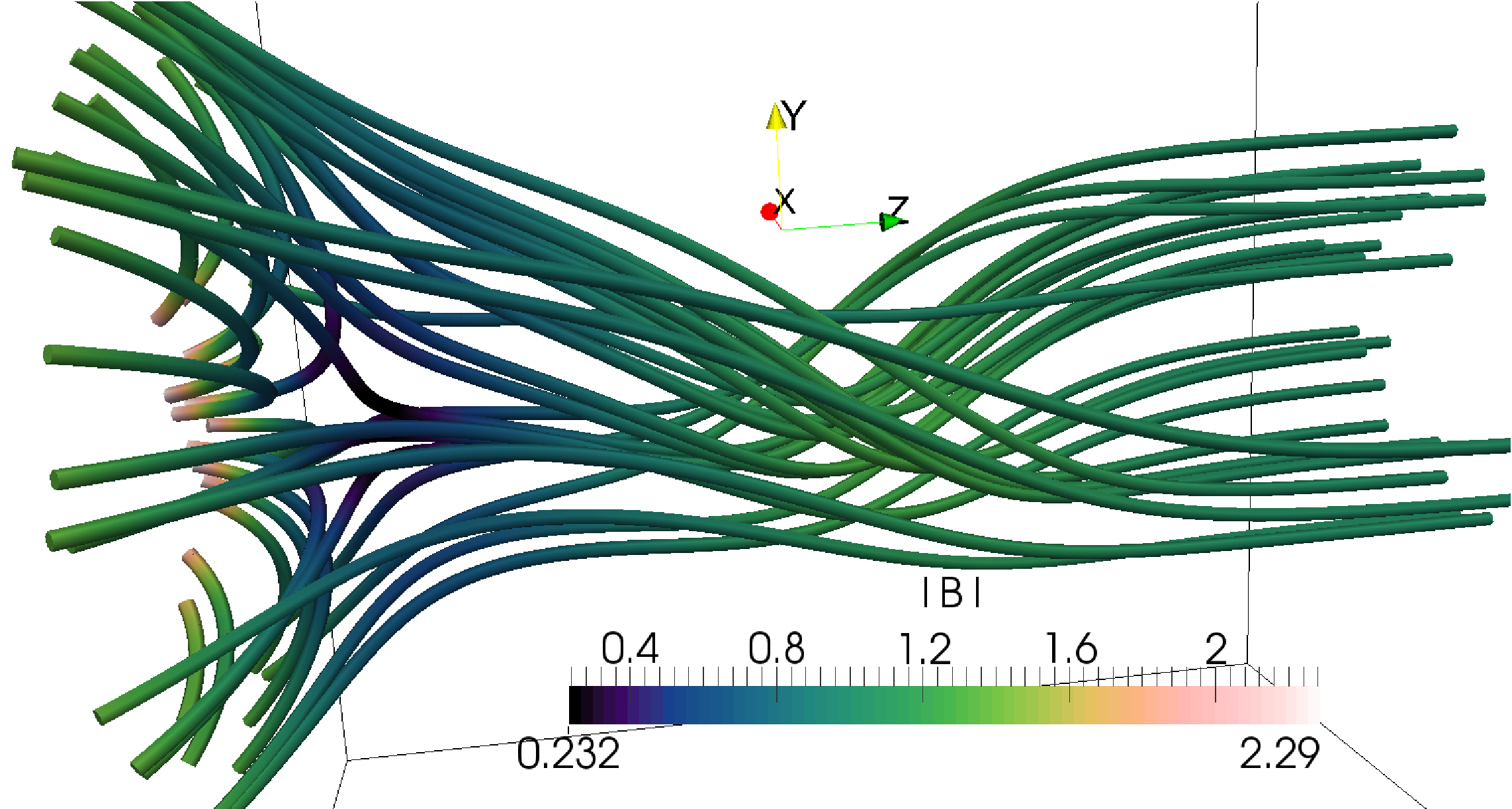}
\end{center}
\caption[]{
Initial magnetic field for the configuration with the magnetic dome
containing the magnetic null and the first twist region of the $E^3$ configuration
at the lower boundary at $z = -8$.
The colors denote the magnetic field strength.
}
\label{fig: e3Domes initB}\end{figure}

With the present framework we are able to investigate the formation of
potentially singular current concentrations around magnetic null-points
where $\BB = 0$.
As noted previously there is strong evidence that in response
to appropriate perturbations singular current concentrations form at nulls in the
perfectly conducting limit
\citep{Syrovatski-1971-33-933-JETP, pontincraig2005,
fuentesfernandez2012, fuentesfernandez2013, Craig-Pontin-2014-788-2-ApJ}.
Here we embed the null point at the base of a coronal loop.
In particular, we take the first twist region of the magnetic field $E^3$
considered in the previous section and
insert a parasitic polarity flux patch on the lower $z$-boundary,
above which is associated a null point within the domain, located at
$(-0.2229, -0.2229, -7.08330)$.
The separatrix surfaces of this null point forms a dome geometry that encloses
the parasitic polarity.
The extent of the domain is from $(-4,-4,-8)$ to $(4,4,0)$ (\Fig{fig: e3Domes initB}).

We study the evolution of this configuration using inertial terms
and velocity damping and replace the evolution of the grid positions by using
equation \eqref{eq: inertial}.
Here we set $\nu = 3$ and choose a grid resolution of $192^3$.
This choice of damping term ensures that the magnetic field does not overshoot
the equilibrium and instead creeps towards it.

\begin{figure}[t!]\begin{center}
\includegraphics[width=0.95\columnwidth]{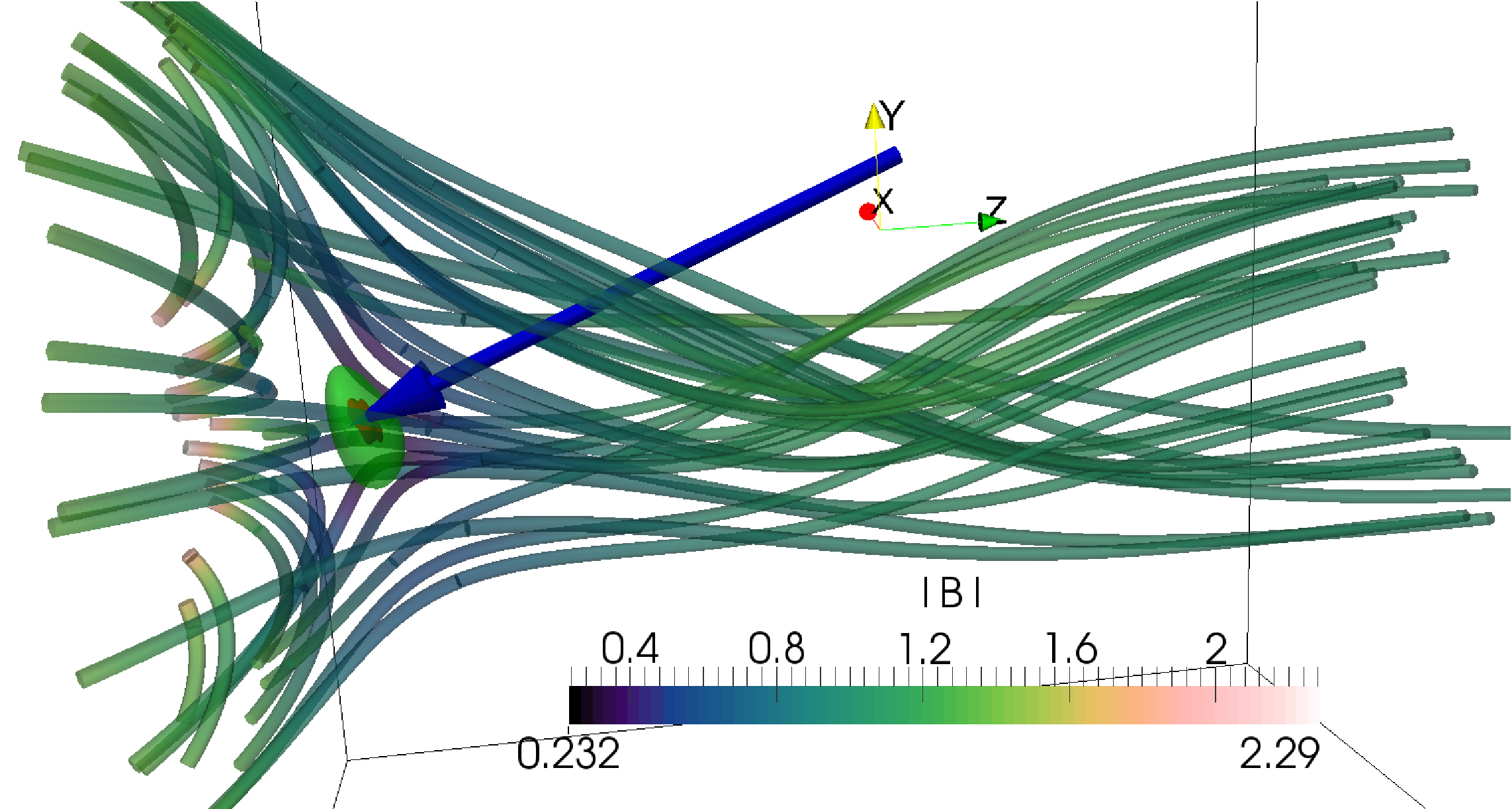}
\end{center}
\caption[]{
Final magnetic field for the first twist region of the $E^3$ configuration
with a magnetic dome together with
isosurfaces of the magnetic field (green, half transparent) and current density
(red opaque).
For the magnetic field we choose a level for the isosurface close to $0$ to
highlight the area around the null, while for $\JJ$ we choose a high value.
It can be seen that the high current concentration lies at the magnetic null.
}
\label{fig: e3Domes finalJ}\end{figure}

As the field evolves it tries to find a relaxed state of reduced Lorentz force.
On average over the domain this does occur.
However, in the absence of plasma pressure,
near the null point the current density increases to such high values that
also the Lorentz force starts to diverge, after which the simulation stops.
The loci of these singular current concentrations are at the magnetic
nulls, as is highlighted in \Fig{fig: e3Domes finalJ}.
In line with previous works this current concentration forms as the spine and
fan of the null point collapse towards one another \citep{pontincraig2005,fuentesfernandez2013}.
To ensure that this is not a numerical artifact one can check that in the absence
of the perturbation -- i.e.\ setting $k = 0$ in Equation (\ref{eq:e3}), there is no
current growth at the nulls.
It should be noted that varying the parameter $\nu$ or resorting to
the magneto frictional approach does not qualitatively change this result.

\begin{figure}[t!]\begin{center}
\includegraphics[width=0.95\columnwidth]{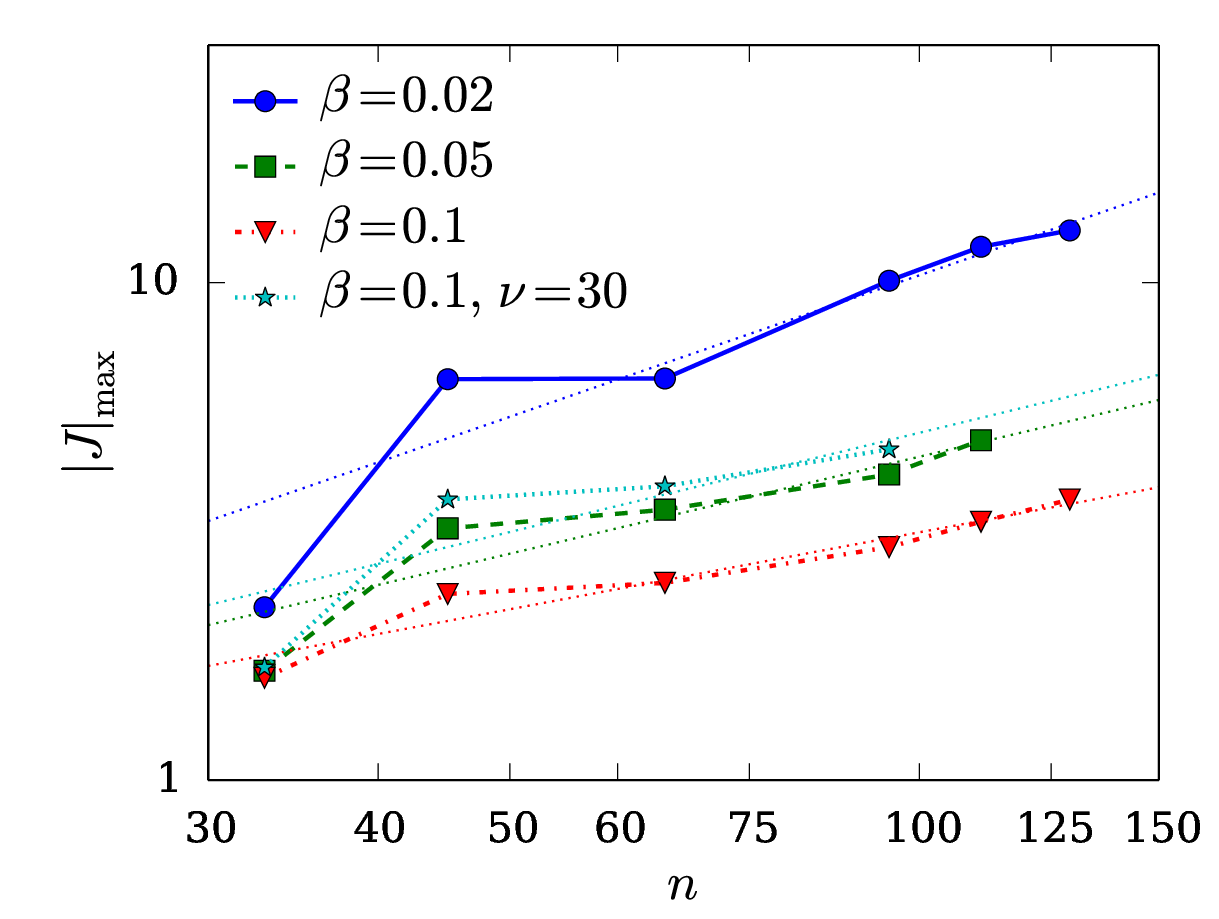}
\end{center}
\caption[]{
Maximum current $|J|_{\rm max}$ at hydrostatic equilibrium for different grid
resolutions and pressure parameters $\beta$ for the configuration with the
magnetic null.
The increase with resolution suggests the existence of a singular current
concentration.
}
\label{fig: e3_d1_mf_gradP_JMax}\end{figure}

Adding a pressure term to our calculations the collapse of the fluid at the
magnetic nulls is halted before the numerical instability sets in.
To achieve this we replace equation \eqref{eq: inertial} by
\eqref{eq: mf gradP} for the evolution of the fluid
and vary the parameter $\beta$ which represents the relative weight of
the pressure gradient to the Lorentz force.
Even with the pressure gradient present we expect singular current concentrations
to form since in general the Lorentz force associated with the null point
collapse is not irrotational, and therefore cannot be balanced by the
pressure gradient \citep{parnell1997,Craig2005,pontincraig2005}.
Indeed, this is what we observe in our simulations where we monitor the
maximum current $|J|_{\rm max}$ in the domain at the stage of
hydrostatic equilibrium (\Fig{fig: e3_d1_mf_gradP_JMax}).
By decreasing $\beta$ the maximum current increases, as the system gets closer
to the zero $\beta$ case.
Increasing the grid resolution we observe a systematic
increase of $|J|_{\rm max}$, suggesting that we
are dealing with a physical current singularity similar to simulations for kink
instability by \cite{Ali-Sneyd-2001-94-221-GwoAstFlDyn}.
This holds also true for the case where we replace the magneto-frictional term
by equation \eqref{eq: inertial}.
As noted by \cite{Craig2005,pontincraig2005},
the effect of the plasma pressure is to weaken the divergent scaling of
the peak current density with resolution, indicating that for large
values of $\beta$ a weaker singularity is present.

\section{Sheared Fields}\label{shearsec}

Past simulations by \cite{Longbottom-Rickard-1998-500-471-ApJ} of sheared
magnetic fields suggested the occurrence of singular current sheets
in the absence of magnetic nulls for sufficiently large shear perturbations.
Such fields would then not reach a smooth force-free equilibrium
supporting the conjecture of \cite{Parker-1972-174-499-ApJ}.
As evidence they pointed to an increasing maximum current density as they
increased the numerical resolution and concluded that the increase will
continue indefinitely.
As maximum resolution they were able to use $65^3$ grid points.

\begin{figure}[t!]\begin{center}
\includegraphics[width=0.95\columnwidth]{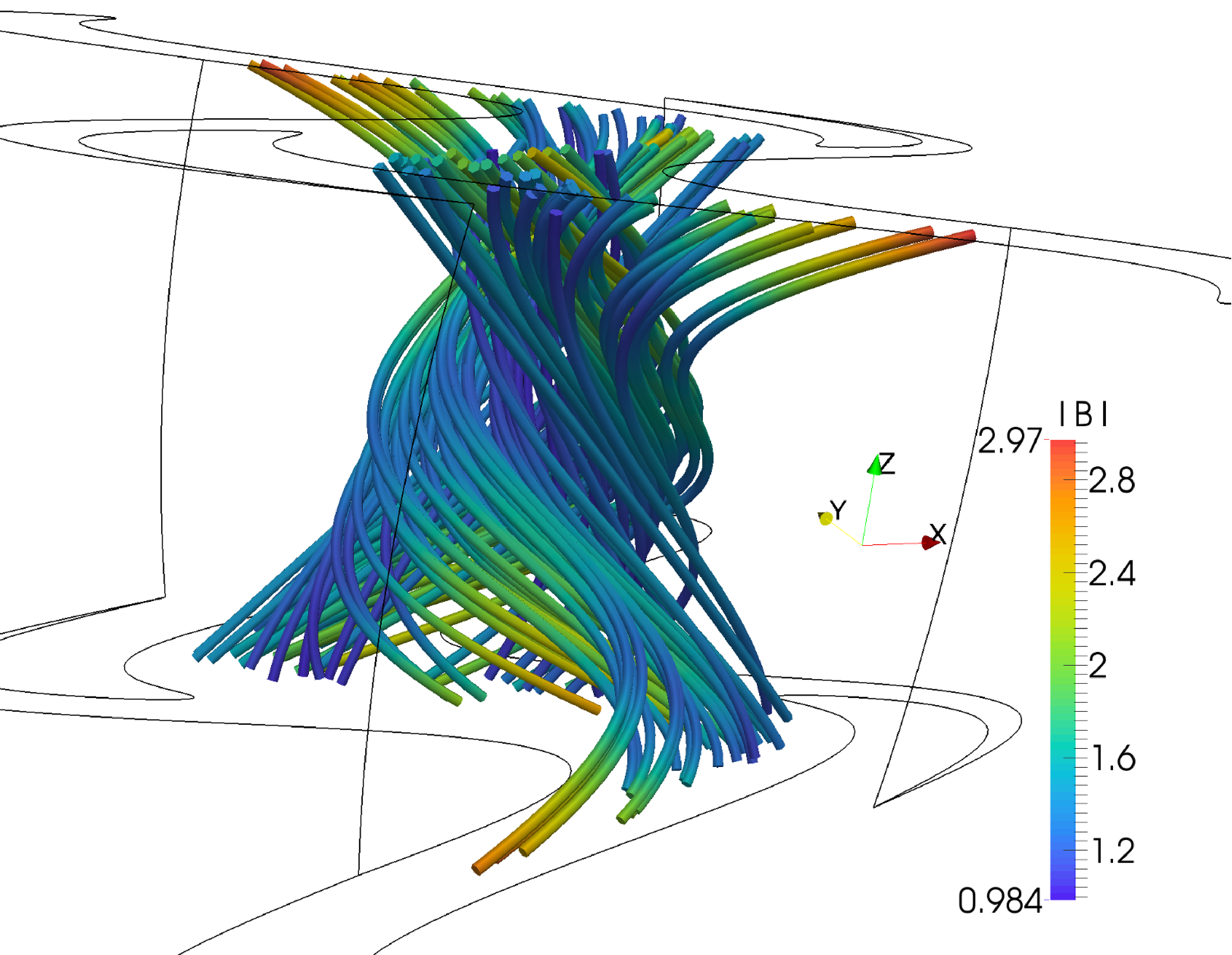}
\end{center}
\caption[]{
Initial magnetic field for the sheared field configuration with $S_B = 1$
together with the distorted grid box.
}
\label{fig: sheared init}\end{figure}

Here we propose that their maximum resolution was too low to make any meaningful
conclusions about the formation of singular current sheets for cases in which
the field is highly sheared.
As remedy we perform simulations with high resolutions and
monitor the formation of current layers.
The field configurations are identical to the ones used by
\cite{Longbottom-Rickard-1998-500-471-ApJ}.
A Cartesian box of size 2 in each dimension is filled with a homogeneous
magnetic field in $z$-direction.
Subsequently the box is distorted in the $y$-direction according to
\EQ
y = y_0-S_A \sin{(2\pi S_K(x_0+O_x)/L_x)}z,
\EN
after which we apply a distortion in the $x$-direction:
\EQ
x = x_0-S_B \sin{(2\pi S_K(y+O_y)/L_y)}z. \label{shear2}
\EN
Here $x_0$ and $y_0$ are the grid coordinates of the undistorted Cartesian grid,
$S_A$ and $S_B$ the shearing strengths, $S_K$ the wave number, $O_x$ and $O_y$
the origins of the coordinate system in $x$ and $y$ and $L_x$ and $L_y$ the length
of the box in $x$ and $y$.
Here we set the size of the undistorted box to $L_x = L_y = L_z = 2$ and center the
domain at the origin.
We choose $S_A = S_K = 1$ in all the runs and vary $S_B$ between $0.1$ and $1$.
Note that the distortion in $x$-direction is performed after the one
in $y$-direction which is why we use $y$ instead of $y_0$ in Equation (\ref{shear2}).
For the $z$-boundaries we apply the line tied condition where the normal component
of the field is fixed and the grid is rigid.
The $x$ and $y$ boundaries are periodic.
An example initial configuration is shown in \Fig{fig: sheared init} for $S_B = 1$.

\begin{figure}[t!]\begin{center}
\includegraphics[width=0.95\columnwidth]{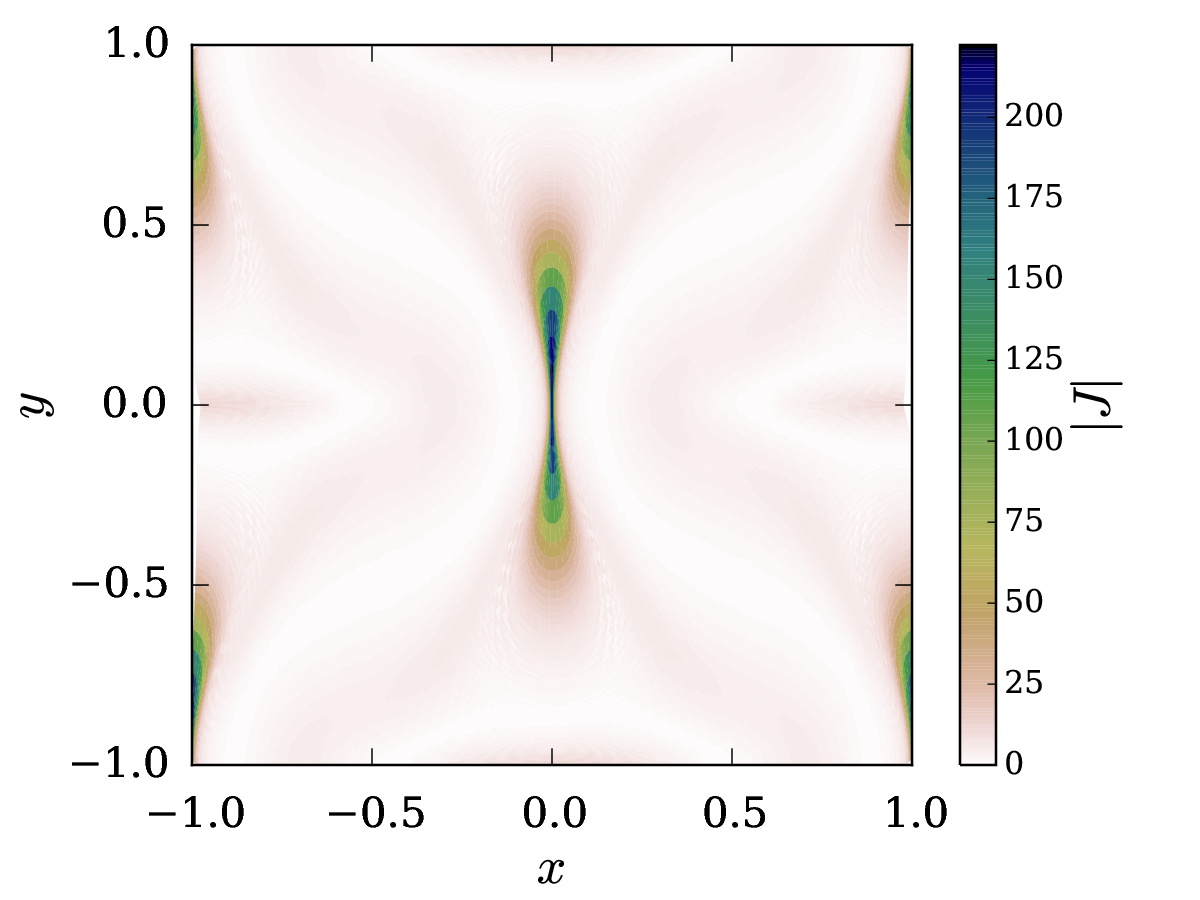}
\end{center}
\caption[]{
Map of the current density for the sheared field numerical experiments
for $S_{B} = 1$ at $z = 0$ and final time.
Although the current concentrates in a small location it is still resolved
with the $180^3$ grid points used here and compressed grid cells at high
current concentrations.
}
\label{fig: st_n180_sb1_Jmap}\end{figure}

\begin{figure}[t!]\begin{center}
\includegraphics[width=0.95\columnwidth]{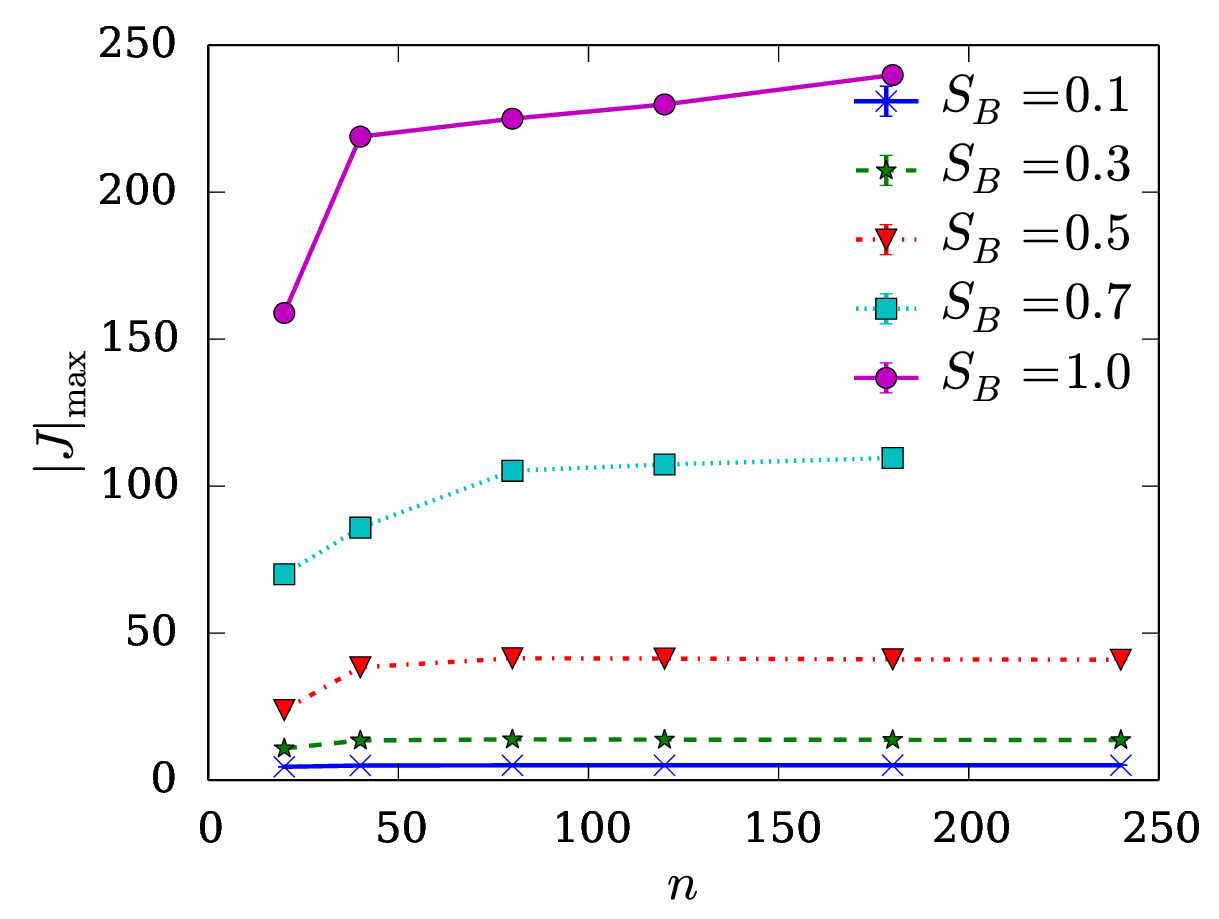}
\end{center}
\caption[]{
Maximum current density $|J|_{\rm max}$ in the saturated state in
dependence of the grid resolution $n$ for various shearing parameters $S_B$
for the sheared field configurations.
For all $S_B$ there is eventually a flattening off of the curves.
}
\label{fig: Jmax}\end{figure}

\begin{figure}[t!]\begin{center}
\includegraphics[width=0.95\columnwidth]{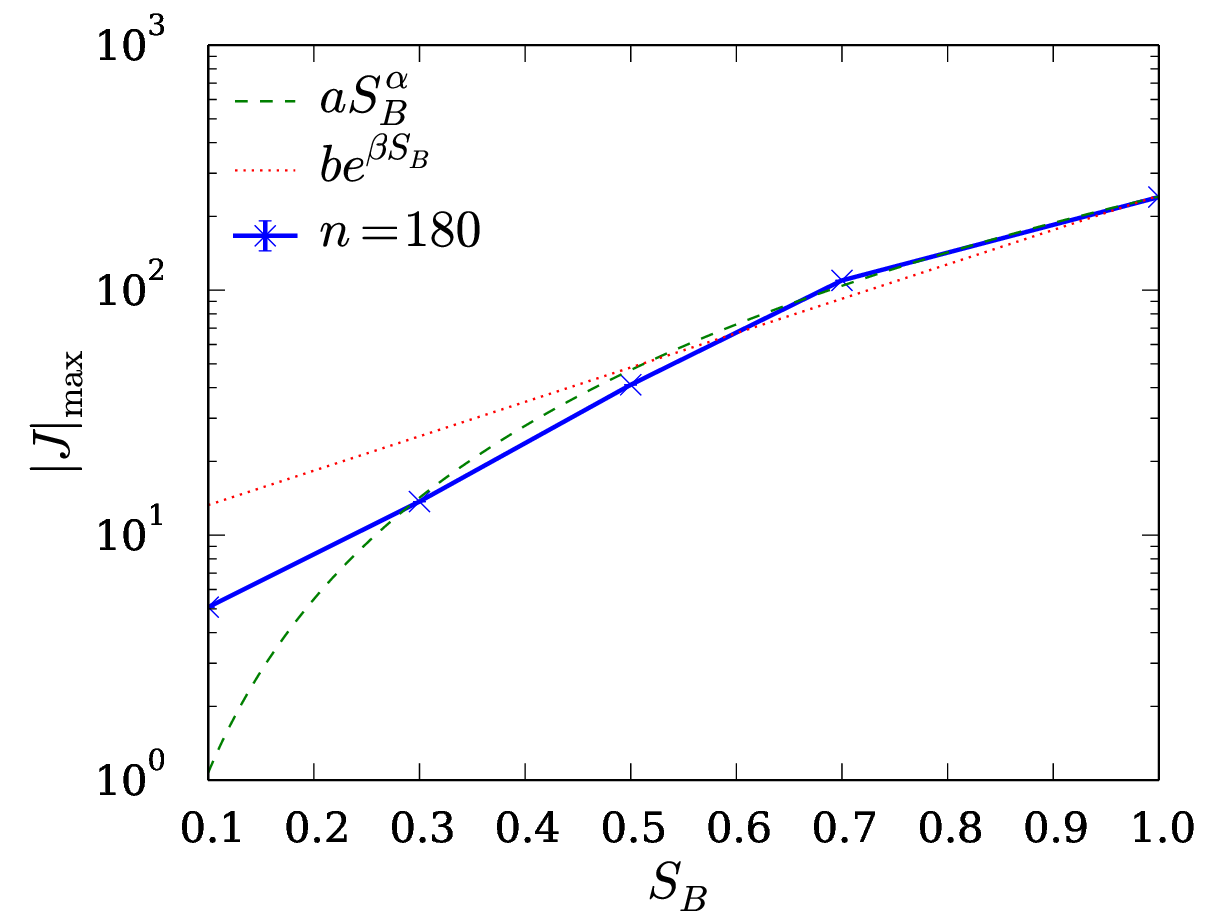}
\end{center}
\caption[]{
Maximum current density $|J|_{\rm max}$ in the saturated state in
dependence of the shear parameter $S_B$ for the resolution $n = 180$
for the sheared field configurations together with a power law and
exponential fit.
Apart from the case with $S_B = 1$ all the values perfectly align with
an exponential law better than with a power law.
For the fit parameters we use a least square method and find
$a = 240.95$, $\alpha = 2.35$, $b = 9.6057$, $\beta = 3.232$.
}
\label{fig: Jmax vs Sb}\end{figure}

As the field relaxes towards a more force-free state, the maximum current in the
simulation domain increases, forming a thin layer running up
the center of the domain, centered on the $z$-axis -- see \Fig{fig: st_n180_sb1_Jmap}.
After some time, however, the growth of the peak current in the domain
flattens off and a stable spatio-temporal maximum $|J|_{\rm max}$ is attained.
Plotting this global maximum of the current as a function of the grid resolution for
large shears we observe an increase with resolution (\Fig{fig: Jmax}), that eventually
saturates -- further increase of the gird resolution does not lead to an increased
current, indicating that an underlying finite current layer has been resolved.
This saturation value strongly increases with shearing parameter $S_B$,
as the field distortion produces strong currents.
\Fig{fig: Jmax vs Sb} shows that this increase is exponential.
This is in line with recent findings by \cite{pontin2015} who found an exponential increase
in the maximum current with increasing twist parameter for the $E^3$ field.
Furthermore, we do not observe any hint for a threshold after which the field shows
current singularities in accordance with Parker's hypothesis.
Already our field with $S_B = 1$ is so strongly twisted that Parker would
have predicted such singularities.

The reason why \cite{Longbottom-Rickard-1998-500-471-ApJ} drew the premature
conclusion that singular current sheets for $S_B > 0.4$
were present was simply due to their limited
maximum resolution, which suggested that above a certain $S_B$ $|J|_{\rm max}$
would grow indefinitely with the resolution suggesting the formation of
singular current sheets.
What is clear is that for the grid resolutions they considered
an unresolved current concentration below the grid scale was present.
However, with our high resolution simulations we are able to resolve the current concentrations
even for high grid distortions (\Fig{fig: st_n180_sb1_Jmap}).

\section{Conclusions}\label{concsec}

We have introduced a new computational code \citep{glemur-github}
that performs an exactly ideal
relaxation towards an equilibrium magnetic field.
This was used to study the
properties of equilibria of various magnetic topologies.
Several implementations of the relaxation procedure were discussed.
We implemented both a magneto-frictional approach and an approach with
velocity damping including plasma inertia.
In both cases a relaxation towards
a force-free state or a magnetohydrostatic equilibrium with finite pressure
were discussed.
The code uses a Lagrangian grid approach, and in contrast
to previous implementations employs mimetic derivatives that lead to an
improved approximation of the final equilibrium \citep{Candelaresi-2014-mimetic}.

We have investigated the ideal evolution of topologically non-trivial magnetic
field configurations and monitored the behavior of the electric current
density.
The emphasis was on determining whether singular current sheets might form
for fields which are sufficiently stressed, as suggested by
\cite{Parker-1972-174-499-ApJ}.
Contrary to Parker's hypothesis
we do not find singular current sheets and all current structures remain
resolved in the absence of magnetic nulls.

The first type of field considered was a braided field that has been previously
well studied.
In support of the previous results
\citep[e.g.][]{Wilmot-Smith-Hornig-2009-696-1339-ApJ}
we find only well resolved current structures.
However, we have noted that at contact areas between regions with different
field line twist, the relaxation of the field towards the force-free state is inhibited,
as measured by various field line integrated quantities.
This suggests that at least
using the artificial path to equilibrium discussed here, there may be a barrier to
reaching the lowest energy state.
This will be discussed further in a future publication.
One should note that, as argued by \cite{pontin2015}, for braided
fields of this nature that exhibit a field line mapping with very small length scales, any
equilibrium that does exist must exhibit current layers on these same small length
scales.
Thus, while Parker's hypothesis for spontaneous formation of current
singularities may not hold for these fields, the proposal that magnetic braiding
can provide a source of coronal heating is still valid.
In particular, as the field is
continually braided by the turbulent convective motions, the length scales of the
current layers will eventually become sufficiently small that reconnection occurs.

We also considered sheared magnetic fields that had previously been implicated
in the formation of current singularities.
We demonstrated that with sufficient grid
resolution, a finite current layer can always be resolved, in contradiction to the
results of \cite{Longbottom-Rickard-1998-500-471-ApJ}, who were severely limited
in the grid resolution available to them.

Lastly, we have considered magnetic fields containing magnetic nulls. We showed
that in their presence, strong and unresolved current structures form
at their loci.
This has been previously observed in various studies
\citep{pontincraig2005,fuentesfernandez2012,fuentesfernandez2013,Craig-Pontin-2014-788-2-ApJ}.
In most of these previous studies a simple linear null point was considered.
Here we considered a coronal loop with a null point near the line-tied boundary
in a separatrix dome configuration -- the perturbation to the field was applied
far form the null point.
Nonetheless, the null point still attracted an intense current.

\acknowledgments

All the authors acknowledge financial support from the UK's
STFC (grant number ST/K000993).
We gratefully acknowledge the support of NVIDIA Corporation with the
donation of one Tesla K40 GPU used for this research.
We are grateful for fruitful discussions with Antonia Wilmot-Smith.

\bibliography{references}

\end{document}